\documentclass[%
reprint, 
superscriptaddress,
amsmath,amssymb,aps,
pra,
floatfix
]{revtex4-2}
\usepackage{caption}
\usepackage{subcaption}
\usepackage{float}
\usepackage{physics}
\usepackage{cleveref}
\usepackage{dsfont}
\usepackage{comment}
\usepackage{amsmath}
\usepackage{mathtools}
\usepackage{xcolor}
\usepackage{graphicx}
\usepackage{dcolumn}
\usepackage{bm}
\usepackage{cleveref}
\usepackage{graphicx}
\usepackage{tabularx}
\usepackage[labelformat=simple]{subcaption}

\begin{document}

\preprint{APS/123-QED}

\title{Quantum memory effects in atomic ensembles coupled to photonic cavities}%

\author{Adam Burgess}
\email{a.d.burgess@surrey.ac.uk}
 \affiliation{Leverhulme Quantum Biology Doctoral Training Centre, University of Surrey, Guildford, GU2 7XH, United Kingdom}
\affiliation{Advanced Technology Institute,  University of Surrey, Guildford, GU2 7XH, United Kingdom}
\affiliation{Department of Physics,  University of Surrey, Guildford, GU2 7XH, United Kingdom}
\author{Marian Florescu}%
\email{m.florescu@surrey.ac.uk}
\affiliation{Advanced Technology Institute,  University of Surrey, Guildford, GU2 7XH, United Kingdom}
\affiliation{Department of Physics, University of Surrey, Guildford, GU2 7XH, United Kingdom}


\date{\today}

\begin{abstract}
In this article we explore the dynamics of many-body atomic systems symmetrically coupled to a single Lorentzian photonic cavity. Our study reveals interesting dynamical characteristics including non-zero steady states, superradiant decay, enhanced energy transfer and the ability to modulate oscillations in the atomic system by tuning environmental degrees of freedom. We also analyse a configuration consisting of a three-atom chain embedded in a photonic cavity. Similarly, we find a strong enhancement of the energy transfer rate between the two ends of the chain and identified specific initial conditions that lead to significantly reduced dissipation between the two atoms at the end of the chain. Another configuration of interest consists of two symmetrical detuned reservoirs with respect to the atomic system. In the single-atom case, we show that it is possible to enhance the decay rate of the system by modulating its reservoir detuning, while in the many-atom case, this results in dynamics akin to the on-resonant cavity. Finally, we examine the validity of rotating wave approximation through a direct comparison against the numerically exact hierarchical equations of motion approach. We find good agreement in the weak coupling regime while in the intermediate coupling regime, we identify qualitative similarities, but the rotating wave approximation becomes less reliable. In the moderate coupling regime, we find deviation of the steady states due to the formation of mixed photon atom states.
\end{abstract}

\maketitle
\section{Introduction}
Recently, there has been a sharp increase in interest in using quantum technologies and their applications to advance present computing techniques. This has primarily been driven by significant improvements in the fabrication of quantum technologies and the robustness of quantum computers with larger numbers of logical qubits, as well as advancements in quantum algorithm designs that have the potential to undermine current cryptographic techniques~\cite{Wang2018,PhysRevA.71.052320,PhysRevA.87.012310,PhysRevA.87.062318,365700}. However, the tendency of quantum systems to decohere, lose quantum correlations, and dissipate energy into the environment due to dephasing and relaxation poses a considerable challenge to their use. For the field of quantum technologies to continue its unprecedented growth in the future, it is essential to fully explore the ways in which the environment affects the evolution of quantum systems and identify ways to control this influence ~\cite{QControl}. Additionally, designing new and improved artificial structures that can process information on shorter spatial and temporal scales has been greatly aided by our growing knowledge of the dynamics of atomic systems in dissipative environments~\cite{CoherentSCQuantumDots,GateOperationQubits,SimulStateMeasureJosephsonQB}. 

In the general theory of open quantum systems\cite{TheoryOQSBook}, spin systems coupled to bosonic reservoirs are an archetypal model used to describe two-level atomic systems interacting with an quantised electromagnetic field.  However, it is exceedingly difficult to unravel the dynamics of the two systems simultaneously; typically, just the dynamics of the atomic system are of interest. To eliminate the environmental degrees of freedom, various approximations are deployed. A standard approximation is the Markov approximation, in which induced memory effects of the environment are neglected. This is appropriate for environments that recover rapidly after interaction with the system~\cite{BornMarkovApproxAtomLaser}.
Although this technique has been successful in analyzing many systems, it fails in adequately capturing quantum-induced memory effects. For instance, the local density of states for the electromagnetic field in micro-structured photonic systems, such as photonic crystals, fluctuates significantly for frequencies close to the photonic band gap edge mode. This invalidates the Markov approximation ~\cite{burgess_modelling_2021,SingleAtomSwitch,john_florescu_2001}, and atomic systems coupled to such a photonic reservoir undergo considerable non-Markovian dynamics. This leads to strong coupling between the atomic system and the photonic reservoir generating dressed atomic states which result in fractional steady state atomic populations, spectral splitting and sub-natural line-widths of the atomic transitions~\cite{John1994}. 

Determining the full impact of structured electromagnetic reservoirs on the dynamics of quantum emitters embedded in them is also crucial as we start to scale up the technology of these artificial systems. It has recently been demonstrated\cite{White2020} that the coherence lifetime of multi-qubit systems may be increased by properly characterising the non-Markovian noise.  Understanding the control mechanisms provided by structured environments may lead to novel architectures for more robust quantum technologies with many applications, including quantum networks~\cite{KimbleQuantumNetwork} and clocks~\cite{QuantumNetworkClocks}. Another major difficulty facing quantum technologies, beyond the preservation of entanglement, is the generation of entanglement. An interesting recent development suggests that it is possible to utilise the non-Markovianity of the environment to generate entanglement between initially uncorrelated atomic systems~\cite{EntangleGenNM,Fleming_2012}. Understanding the bi-directional flow of information between system and environment - the essence of non-Markovian systems - makes it possible to transfer information and, therefore, quantum correlations between systems using the environment as an intermediary.  Photonic cavity systems are naturally suited for generating such inter-atomic entanglement, as the atoms couple strongly to the central cavity mode allowing for efficient energy transfer between atoms, even without direct inter-atomic coupling. Here we intend to study such systems.

This article is structured as follows: Section II covers the theoretical formalism of for describing $N$ atomic systems symmetrically coupled to one another inside a photonic reservoir associated with Lorentzian cavity and explores the influence of the reservoir characteristics on the induced system of interest dynamics. In Section III, we explore a configuration of a three-atom chain coupled via nearest neighbour interactions inside a Lorentzian cavity and identify target initial conditions that can lead to significantly reduced dissipation. In Section IV, we consider a many-atom systems coupled to two symmetrically detuned cavities and investigate the way in which the structure of the reservoir impacts the decay rate of the atomic system excitations. Finally, in Section V, we provide a direct comparison of the results obtained under the RWA approximation against the predictions of the numerically exact hierarchical equations of motion approach. 

\section{Many Atoms in a Photonic Cavity}
\begin{figure}
    \centering
    \includegraphics[width=0.4\textwidth]{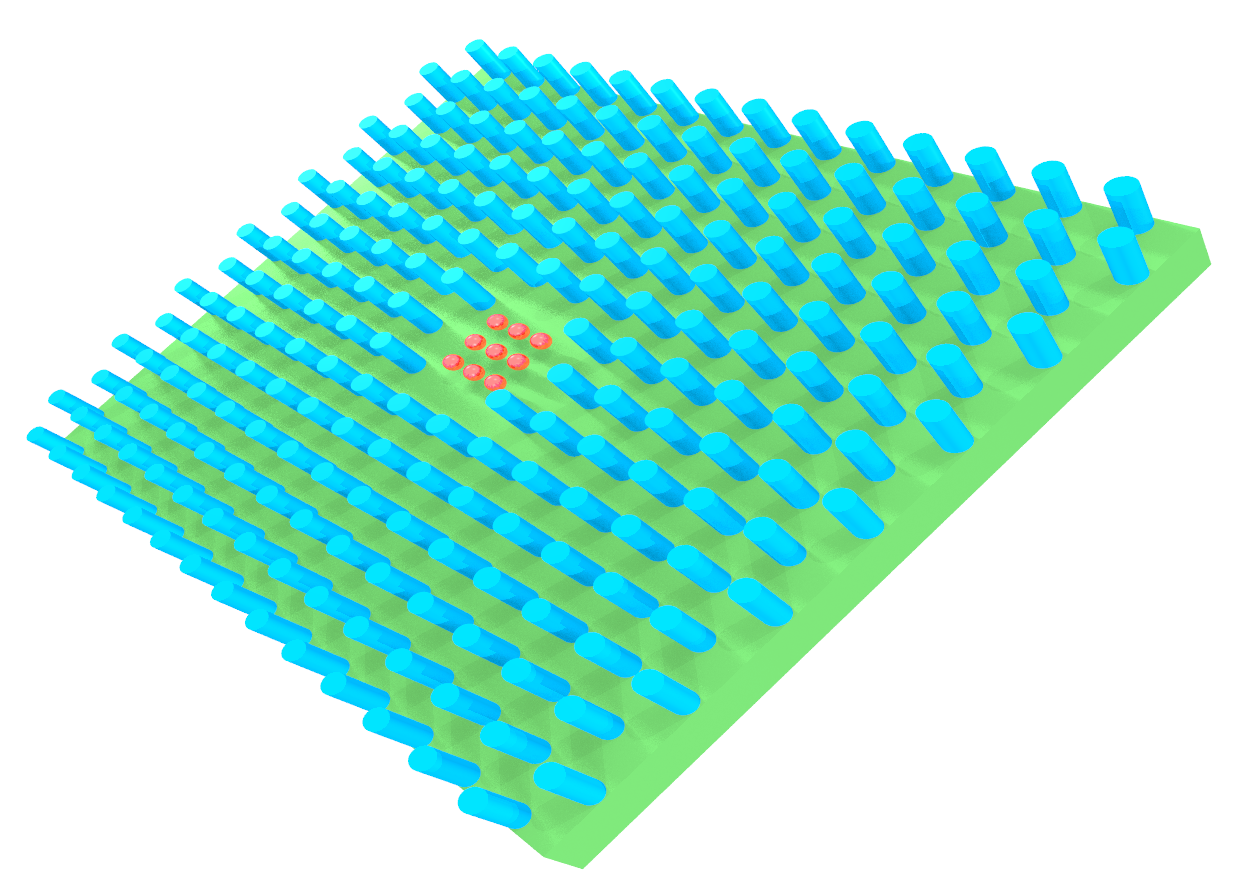}
    \caption{A schematic of many atoms embedded within a photonic cavity reservoir.}
    \label{fig:Cavitypic}
\end{figure}

\begin{figure*}[ht]
    \centering
    \begin{subfigure}[b]{0.4\textwidth}
    \centering
    \includegraphics[width=\textwidth]{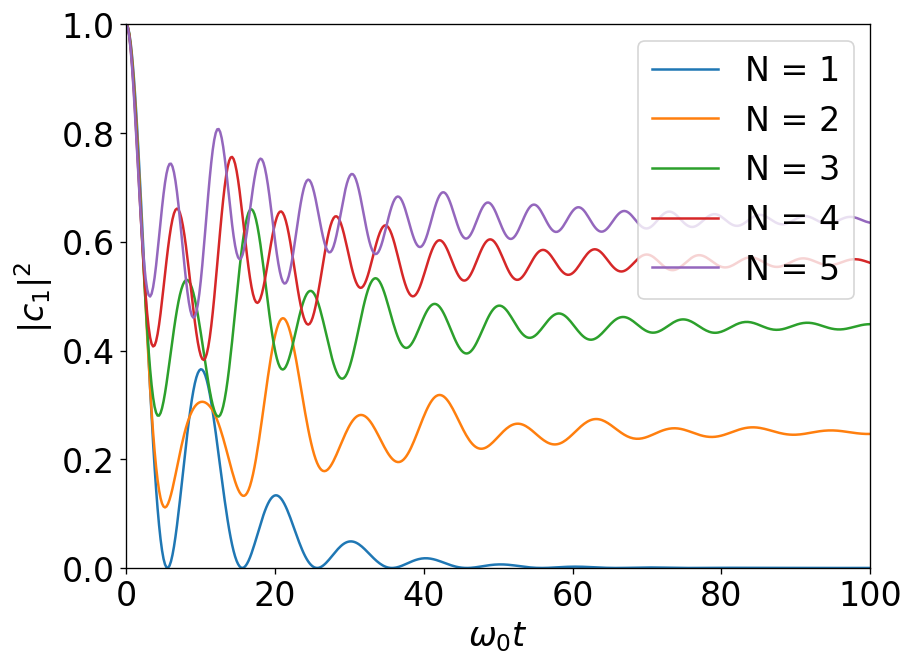}
    \caption{ }
    \label{fig:NAtomsSingleExcited}
\end{subfigure}
\begin{subfigure}[b]{0.4\textwidth}
    \centering
    \includegraphics[width=\textwidth]{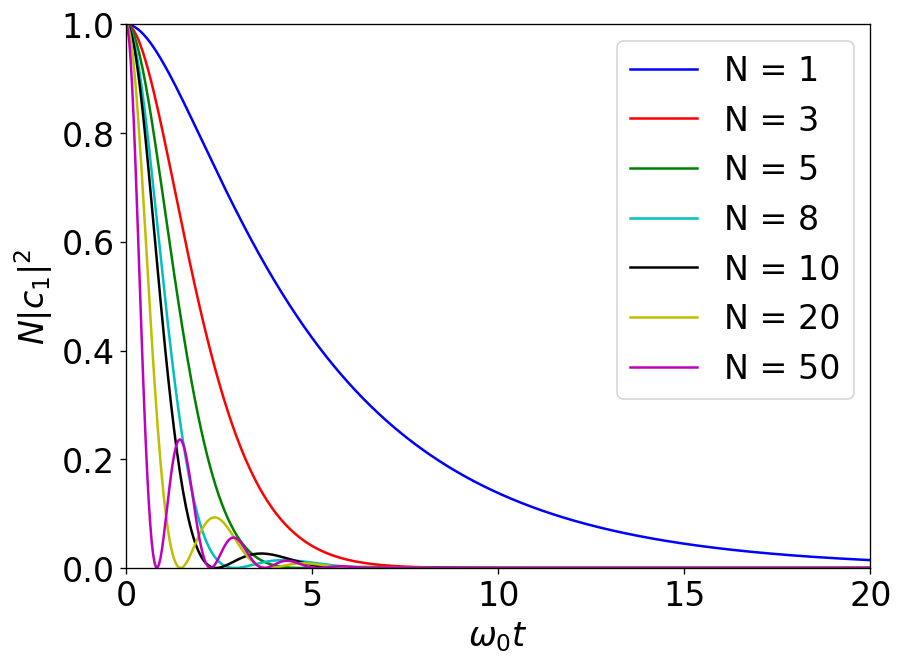}
    \caption{ }
    \label{fig:SuperRadient}
\end{subfigure}
\\
    \begin{subfigure}[b]{0.4\textwidth}
    \centering
    \includegraphics[width=\textwidth]{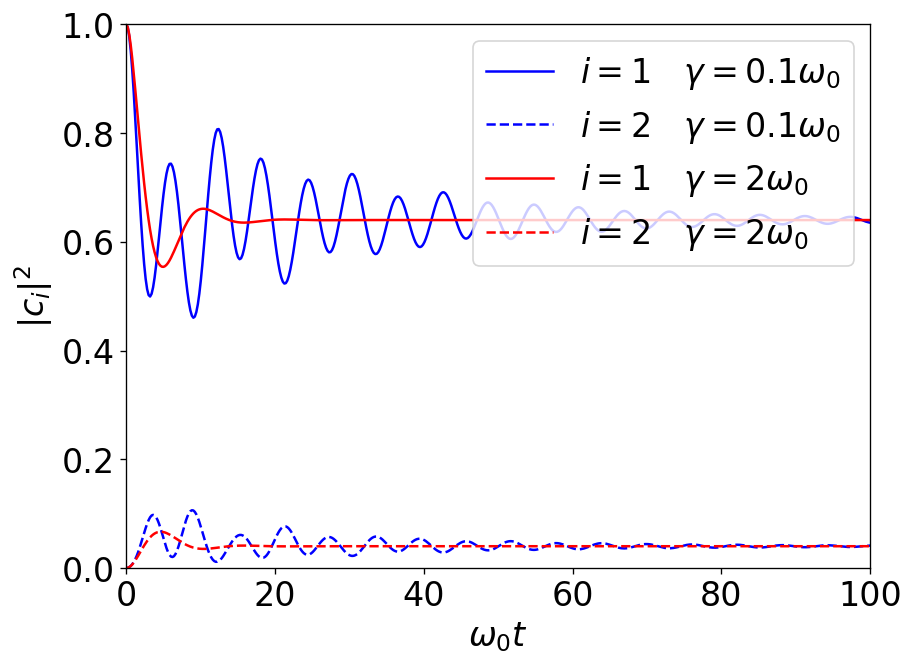}
    \caption{ }
    \label{fig:FCEnhancement}
\end{subfigure}
\begin{subfigure}[b]{0.4\textwidth}
    \centering
    \includegraphics[width=\textwidth]{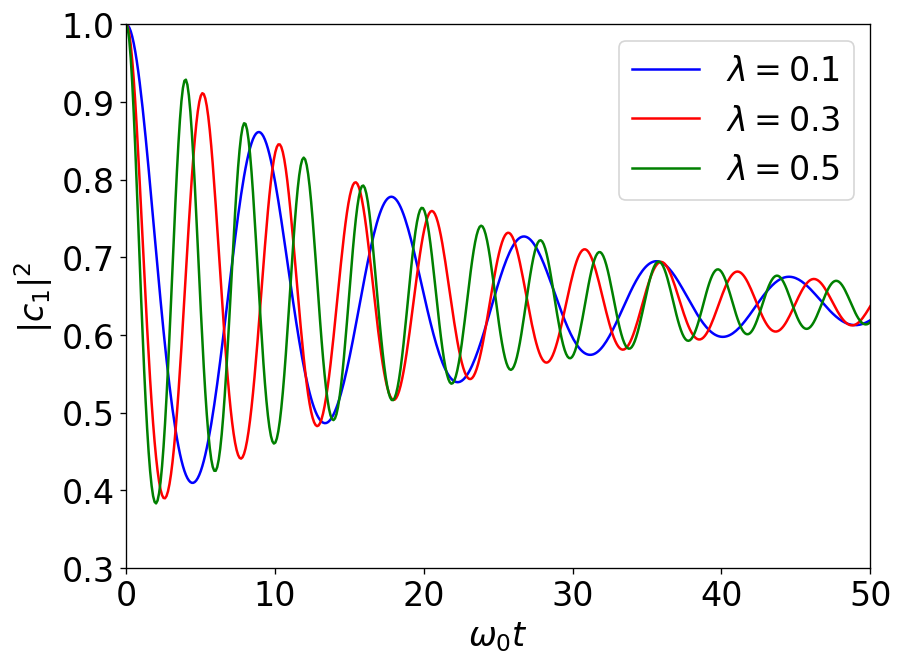}
    \caption{}
    \label{fig:Tuneable}
\end{subfigure}
    
    \caption{(a) A plot of the time evolution of the excited state population of a single initially excited atom coupled to varying numbers of atomic systems with $\Lambda=\gamma=\lambda=0.1\omega_0$. (b) Enhancement of dissipation rate with the number of coupled atomic systems. A plot of the time evolution of the excited state population of a single initially excited atom coupled to a varying number of atoms $N-1$. $\Lambda=0,\lambda=0.1\omega_0, \gamma=\omega_0$. (c) Enhancement of energy transfer rate with the quality of the cavity. A plot of the time evolution of the excited state population of a single initially excited atom coupled to 4 other atoms for varying spectral width $\gamma$. $\Lambda=\lambda=0.1\omega_0$. (d) Tuneable frequency of oscillations by modulation of the coupling strength $\lambda$ for five coupled atoms in the Lorentzian cavity.   $\Lambda=0,\gamma=0.1\omega_0$.  }
    \label{fig:my_label}
\end{figure*}

We begin our analysis by considering $N$-atoms embedded within a photonic cavity. A schematic architecture of this system is shown in Fig.~\ref{fig:Cavitypic}. The atoms couple to the local electromagnetic field through the interaction of their transition dipole with the electric field associated with the random fluctuations in the environment and there is also a direct inter-atomic coupling generated by dipole-dipole interaction. The Hamiltonian associated with this model is given by 
\begin{align}
    H = &\sum_{i=1}^N\omega_0\sigma_i^+\sigma_i^- + \sum_\lambda \omega_\lambda a^\dag_\lambda a_\lambda +  \nonumber \\&i\sum_{i,\lambda}\sigma_i^xg_\lambda(a_\lambda -a^\dag_\lambda ) + \sum_{i\neq j}\Lambda\sigma^+_i\sigma_j^-,
\end{align}
where $\sigma_i^+$and $\sigma_i^-$, are the excitation and de-excitation operators for the $i^\textrm{th}$ atomic system respectively, $a_{\lambda}$ and $a^{\dagger}_{\lambda}$ are the bosonic field annihilation and creation operators, $\omega_0$ and $\omega_{\lambda}$ are the atomic transition and the $\lambda$-boson mode frequencies,  $g_\lambda$ is the coupling strength of the atomic system and the $\lambda$-boson mode~\cite{Pfeifer,AllenEberly,SuperRadiancePBG} and $\Lambda$  is the dipole-dipole coupling strength between atoms.
For systems that interact with a local-electromagnetic field that is near resonant with the systems' transition energies, it is useful, and effective, to deploy the rotating wave approximation (RWA). This allows for the neglecting of terms that do not conserve the number of excitations in the system as they contribute rapidly oscillating terms into the dynamics and have a vanishing average.  The RWA Hamiltonian is then given by
\begin{align}
    H = &\sum_{i=1}^N\omega_0\sigma_i^+\sigma_i^- + \sum_\lambda \omega_\lambda a^\dag_\lambda a_\lambda +  \nonumber \\&i\sum_{i,\lambda}g_\lambda(a_\lambda \sigma^+_i -a^\dag_\lambda \sigma^-_i ) + \sum_{i\neq j}\Lambda\sigma^+_i\sigma_j^-
    \label{eqn:RWA}
\end{align}
 This is equivalent to a Dicke model~\cite{Dicke} for $N$-atoms with a coupling between each of the atoms and is an archetypal model used in quantum optics. We note here that on a practical side, the RWA allows for the generation of analytical solutions to many quantum optical problems.

We have shown previously~\cite{BurgessSingleEx} that is possible to explore the single-excitation dynamics of any system governed by the Hamiltonian above by using the he environment correlation function and the associated characteristic  spectral density. The spectral density simply defines the distribution and coupling strengths of our electromagnetic field modes,
\begin{equation}
    J(\omega) = \sum g_\lambda^2 \delta(\omega-\omega_\lambda).
\end{equation}

For a lossy cavity, the spectral density is given by the Lorentzian distribution. 

\begin{equation}
    J(\omega) = \frac{\lambda\gamma}{\gamma^2+(\omega-\omega_0)^2},
    \label{eqn:specD}
\end{equation}
where $\gamma$ represents the spectral width of the Lorentzian distribution and, thus, provides a measure of how dissipative the cavity is. $\lambda$ defines the coupling strength of the cavity to the atomic system of interest. 

If we consider the single excitation wavefunction given by 
\begin{equation}
    \phi(0) = \sum_i^N c_i(0)\psi_i +\sum_\lambda c_\lambda(0) \psi_\lambda,
\end{equation}
where $\psi_i = \ket{i}_A\ket{0}_B$ is the state wherein the $i^\textrm{th}$ atom is in its excited state, and all other atoms and the reservoir are in their ground states. $\psi_\lambda = \ket{0}_A\ket{\lambda}_B$ represents the quantum states with all atoms in their ground state, and the photon reservoir with its $\lambda$ mode excited.
The time evolution of this state is given by
\begin{equation}
    \phi(t) = c_0\psi_0 + \sum_i^N c_i(t)\psi_i +\sum_\lambda c_\lambda(t) \psi_\lambda,
\end{equation}
due to conservation of excitation number. 
It is convenient for our discussion to consider the total polarisation of the system, $c_+(t)= \sum_i^N c_i(t)$.
By adiabatic cancelling of the reservoir modes we generate the following equations of motion 
\begin{align}
    \Dot{c}_i =& -iJ(c_+-c_i) - \int^t_0G(t-t_1)c_+(t_1)dt_1,\nonumber \\
    \Dot{c}_+ =& -iJ(N-1)c_+ - N\int^t_0G(t-t_1)c_+(t_1)dt_1. 
    \label{eqn:EOM}
\end{align}
We have introduced here the memory kernel
\begin{equation}
    G(t) = \sum_\lambda g_\lambda^2 e^{i(\omega_0-\omega_\lambda)t}.
\end{equation} 
For the Lorentzian cavity system described by the spectral density \ref{eqn:specD}, this takes the form
\begin{equation}
    G(t) = \lambda e^{-\gamma t}.
\end{equation}
By utilising the Laplace transform, we can solve for the time evolution of the $i^\textrm{th}$ atoms excited state amplitude $c_i$, as derived in Appendix \ref{app:FC}, yielding 
\begin{gather}
c_i(t)= (c_i(0) -\frac{c_+(0)}{N})e^{i\Lambda t} +\\ \frac{c_+(0)}{N}e^{-\mu^* t}\left(\cosh(\Gamma t) + \frac{\mu}{\Gamma}\sinh(\Gamma t)  \right)\nonumber
\end{gather}
where
\begin{equation}
    \Gamma =\sqrt{ \mu^2-\lambda N},
\end{equation}
and 
\begin{equation}
    \mu = \frac{1}{2}\big(\gamma-i\Lambda(N-1)\big).
\end{equation}

We note  that the effects of the environment are entirely determined by the total polarisation of the system $c_+$ due to the symmetric nature of the atomic couplings. 
In the long-time limit $t\rightarrow\infty$ the populations become
\begin{equation}
    P_{i\infty} = \lim_{t\rightarrow\infty} |c_i(t)|^2 = |c_i(0)-\frac{c_+(0)}{N}|^2.
\end{equation}

Our results shown Fig.~\ref{fig:NAtomsSingleExcited} unveil the influence of  size pf the atomic system ensemble on the dynamics of the atomic populations. Clearly, the steady state of $(1-1/N)^2$ is attained for initially excited atom 1 $c_1(0)=c_+(0)=1$. We note that due to the factor $\Gamma$ in the $\cosh$ and $\sinh$ functions, for any value of $N>1$, the populations of the atomic systems will undergo oscillatory evolution, as long as the initial total polarisation is non-zero ($c_+(0)\neq 0$). If we turn off the dipole-dipole coupling ($\Lambda=0$), the system displays oscillatory behaviour only when $\gamma<4\lambda N$. This shows that we can generate oscillatory behaviour by increasing the number of atomic systems $N$ coupled to our reservoir. However, by reducing the real component of $\Gamma$ we reduce the counter-action to the exponential decay, increasing the effective decay rate in the system and leading to superradiant decay for short times and some transient resurgence due to the non-Markovian effects (see Fig.~\ref{fig:SuperRadient}) . We note that the exponential decay rates scale with the number of atoms in the systems. To see this more explicitly, we  consider the Taylor expansion around $t=0$ up to the fourth order. Assuming that $c_i(0) = \dfrac{1}{\sqrt{N}}$ then 
\begin{align}
    P_i(t) &= |c_i(t)|^2 =\\ &\frac{1}{N}\left(1 - \lambda N t^2 + \frac{2}{3}\lambda\mu N t^3 - \frac{1}{3}\lambda N (\mu^2-\lambda N) t^4  \right). \nonumber
\end{align}

However, this superradiance is limited, as the dynamics are still enveloped by the decay from $e^{-\mu^*t}$ as in Fig.\ref{fig:SuperRadient}, the decay increases with atomic system number, but this saturates for large values of $N$, it appears to increase, however, this is due to an introduced oscillation in the system but the enveloping decay is the same.
Interestingly, we also note that as the spectral width of the cavity $\gamma$ decreases, there is an increase in the effective transfer rate between the atomic systems as 
\begin{equation}
    \lim_{\gamma\rightarrow 0} \Gamma = i\sqrt{\frac{1}{4}\Lambda^2(N-1)^2 + \lambda N}
\end{equation}
which implies that the coupling to the cavity mode increases the energy transfer rate between the atomic systems. This has been shown in Fig.~\ref{fig:FCEnhancement}, wherein by modulating the spectral width of the reservoir $\gamma$, where is apparent that the excitation is transferred at a faster rate when $\gamma$ takes smaller values. This suggests that the environment supports energy transfer and provides an additional channel. Furthermore, we can tune this rate for small values of $\gamma$ by modulating the coupling strength to the cavity mode $\lambda$. As shown in Fig.~\ref{fig:Tuneable}, we can now infer some control over the dynamics of our system by modulating the environment.

\section{Atomic Chain in Lorentzian Cavity}
\begin{figure*}[ht!]
    \centering
    \begin{subfigure}[b]{0.3\textwidth}
    \centering
    \includegraphics[width=\textwidth]{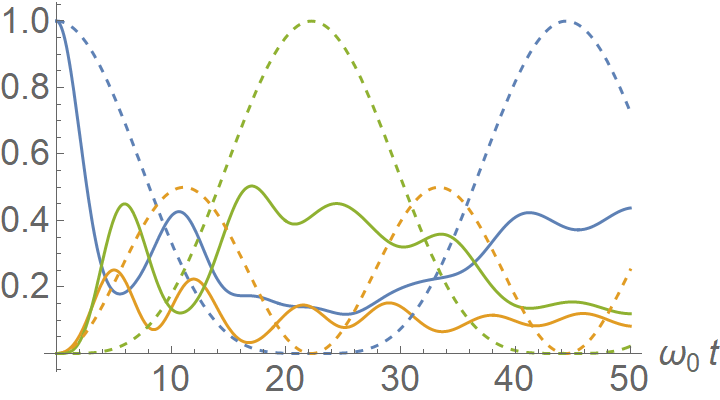}
    \caption{ }
    \label{fig:3A}
\end{subfigure}
\begin{subfigure}[b]{0.3\textwidth}
    \centering
    \includegraphics[width=\textwidth]{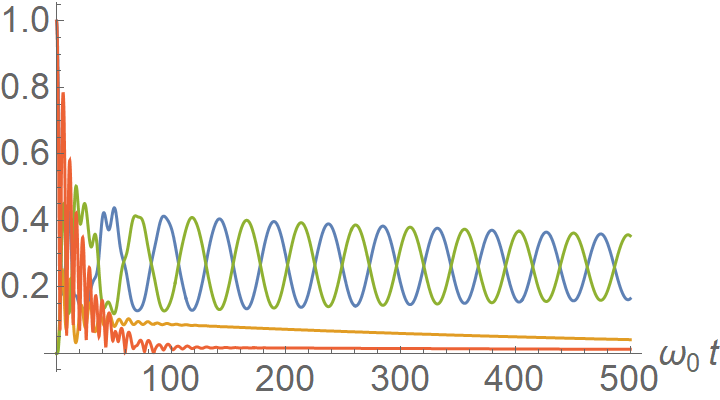}
    \caption{ }
    \label{fig:3ALong}
\end{subfigure}
\begin{subfigure}[b]{0.3\textwidth}
    \centering
    \includegraphics[width=\textwidth]{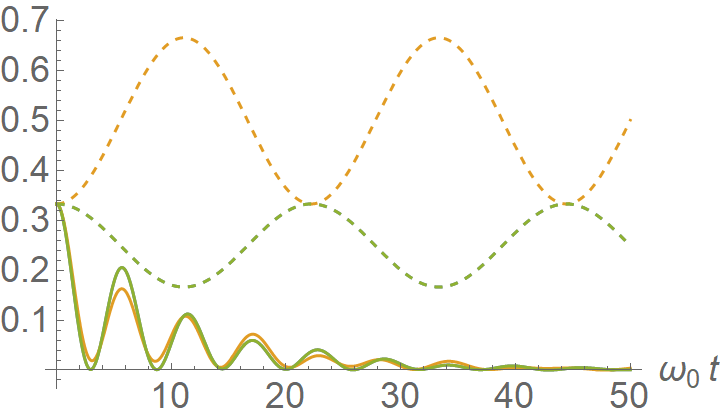}
    \caption{ }
    \label{fig:3AEq}
\end{subfigure}
\\
\begin{subfigure}[b]{0.3\textwidth}
    \centering
    \includegraphics[width=\textwidth]{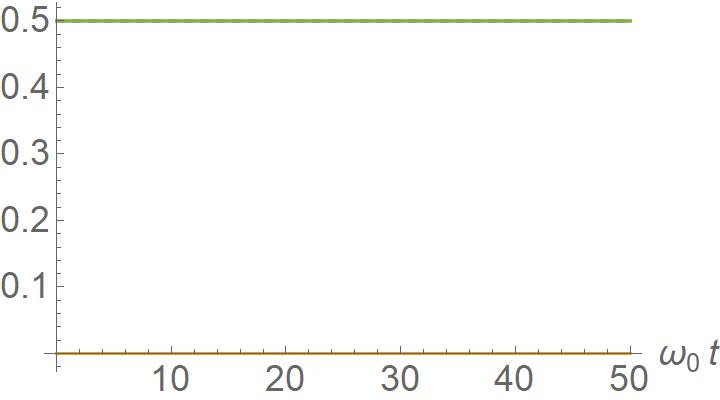}
    \caption{}
    \label{fig:3ANegE}
\end{subfigure}
\begin{subfigure}[b]{0.3\textwidth}
    \centering
    \includegraphics[width=\textwidth]{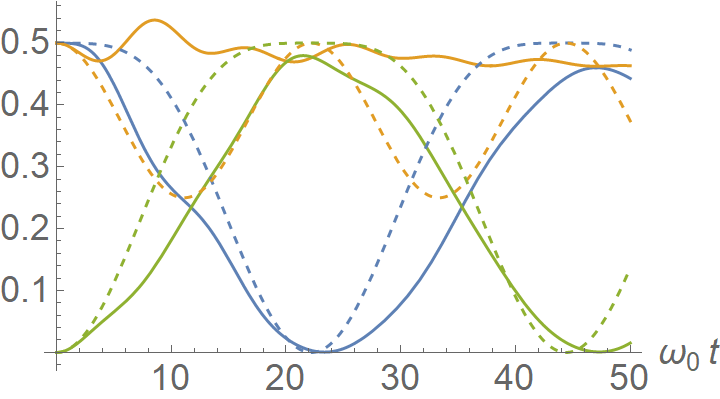}
    \caption{}
    \label{fig:3AAsym0}
\end{subfigure}
 \begin{subfigure}[b]{0.3\textwidth}
    \centering
    \includegraphics[width=\textwidth]{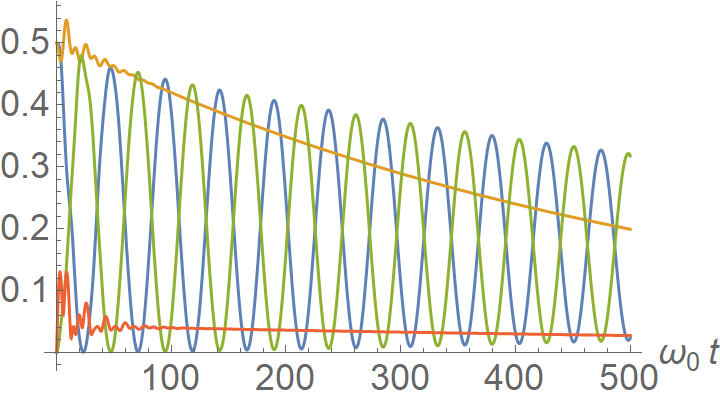}
    \caption{}
    \label{fig:3AAsym0Long}
\end{subfigure}   
    \caption{Plots of the time evolution of the excited state population of the three atoms in a chain in a photonic cavity (thick) and free (dashed) for the first atom (blue), second (orange), third (green).  (a) When the first atom is initially excited. (b) The same as (a) but extended up to late times and containing the total polarisation $|c_+| = |c_1+c_2+c_3|$ (dark orange). (c) When all atoms start with the same polarisation $c_i(0) = \frac{1}{\sqrt{3}}$. (d) When both the atoms at the ends of the chain are initialised with an equal magnitude but opposite sign $c_1(0)=-c_3(0)=\frac{1}{\sqrt{2}}$. (e) When both the first two atoms of the chain are initialised with an equal magnitude but opposite sign $c_1(0)=-c_2(0)=\frac{1}{\sqrt{2}}$. (f) as in (e) but extended to later times, with total polarisation $|c_+|$ (dark orange). $\Lambda=0.1\omega_0$, $\gamma =0.1\omega_0$ and $\lambda=0.1\omega_0$.  }
    \label{fig:3Aall}
\end{figure*}
In the previous section, we considered the idealised system in which we assumed symmetric all to all coupling between the atomic systems. A system of this kind are realisable, but some creative engineering of the physical platform is required, for example, employing coupled Josephson parametric oscillators and cavity 
buses\cite{Majer2007,Onodera2020}. A slightly more practical physical instantiation would consist of atomic chains embedded in cavity structures, such that the atomic systems are coupled to their nearest neighbours. Here, we consider the dynamics of a three-atom chain in a photonic cavity.

For a chain of atomic systems of three atomic systems embedded within a single photonic cavity, we system evolution is described by the following Hamiltonian
\begin{align}
        H = &\sum_{i=1}^3\omega_0\sigma_i^+\sigma_i^- + \sum_\lambda \omega_\lambda a^\dag_\lambda a_\lambda +   \\&i\sum_{i,\lambda}g_\lambda(a_\lambda \sigma^+_i -a^\dag_\lambda \sigma^-_i ) + \sum^3_{i}\Lambda\big(\sigma^+_i\sigma_{i+1}^- + \sigma^+_{i+1}\sigma_{i}^-\big)\nonumber.
\end{align}
where we are using the same notations as in Eqn.~\ref{eqn:RWA}, but we note that now the atoms only interact with their nearest neighbour.

Following the same procedure as before we obtain the following equations of motion for the excited state amplitudes 
\begin{align}
    \Dot{c}_1 &= -i\Lambda c_2 -\int_0^t G(t-t_1)c_+(t_1) dt_1 \nonumber\\
    \Dot{c}_2 &= -i\Lambda (c_+-c_2) -\int_0^t G(t-t_1)c_+(t_1) dt_1 \nonumber\\
    \Dot{c}_3 &= -i\Lambda c_2 -\int_0^t G(t-t_1)c_+(t_1) dt_1 .
    \label{eqn:chain}
\end{align}

These coupled differential equations in Eqn.~\ref{eqn:chain} can be solved through Laplace transformations. Before that, we can identify a few dynamical qualities by looking at the form of these differential equations. If we turn off the inter-atomic coupling $\Lambda=0$, we are led back to the previous system studied. Let's consider initial conditions where $c_1(0)=-c_3(0)$ and that $c_2(0)=0$. In this case, there is no time evolution for the system as the total polarisation of the atoms seen by the bath and by the central atom is zero.
We have demonstrated this for various initial conditions in Fig. \ref{fig:3Aall}. In Fig.~\ref{fig:3A}, we consider that the first atom in the chain is initially excited and note that the rate of energy transfer between the first and third atoms becomes higher when the cavity is introduced. This is because the cavity acts as a secondary channel for the excitation to pass between atoms, supplementing the conventional hopping along the chain. We also note that the system achieves a lower occupation of the excited state; this is natural as the cavity is dissipative, so we anticipate the loss in excitation. Furthermore, as shown Fig.~\ref{fig:3ALong}, at longer times, the effects of the cavity become weaker and weaker, leading to free-like oscillations between the first and third atoms in the system. Physically, this can be interpreted as follows: the total polarisation $c_+$, which couples the atoms to the cavity, has decayed away and  the cavity can no longer interact these atomic systems. This a remarkable effect as it provides direct means to effectively decouple these atomic systems from their local photonic environment. When we prepare the atomic systems with the same initial amplitudes ($c_i = \frac{1}{\sqrt{3}}$) as shown in Fig.~\ref{fig:3AEq}, the evolution of the atomic systems is akin to the single atom case with slight modulations for the second atom as its coupling is different to the two end atoms. By choosing the two atoms at the end of the chain to have equal but opposite initial amplitudes ($c_1(0)=-c_3(0)=\frac{1}{\sqrt{2}}$), as shown in Fig.~\ref{fig:3ANegE}, we have no evolution as predicted. 
Finally, in Fig.~\ref{fig:3AAsym0} and \ref{fig:3AAsym0Long}, the first two atoms in the chain are prepared with initial amplitudes of equal magnitude but the opposite sign ($c_1(0)=-c_2(0)=\frac{1}{\sqrt{2}}$).
 Here the total polarisation $c_+$ is initially zero; as such, we do not anticipate much dissipation, though this will grow as the excitation is passed towards the third atom leading to some dissipation. Interestingly, in Fig.~\ref{fig:3AAsym0Long} we note that as the system evolves towards later times, the amount of dissipation is, in fact, incredibly low, and the two end atomic systems act as if they were nearly free systems. Furthermore, their effective detuning is considerably lower than in the case of Fig.~\ref{fig:3ALong}, suggesting this would be a more useful scheme to generating these free dynamics. Clearly, the results presented here demonstrate that despite a lossy environment, it is still possible maintain control over the system evolution, and to massively reduce the amount of dissipation while retaining the energy transfer between atomic systems. 

\section{Double Dissipative cavity}
\begin{figure*}[ht!]
    \centering
     \begin{subfigure}[b]{0.3\textwidth}
    \centering
    \includegraphics[width=\textwidth]{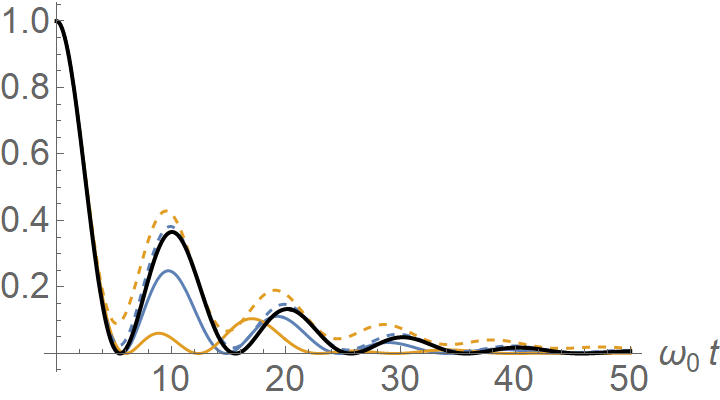}
    \caption{ }
    \label{fig:2CvC}
\end{subfigure}
\begin{subfigure}[b]{0.3\textwidth}
    \centering
    \includegraphics[width=\textwidth]{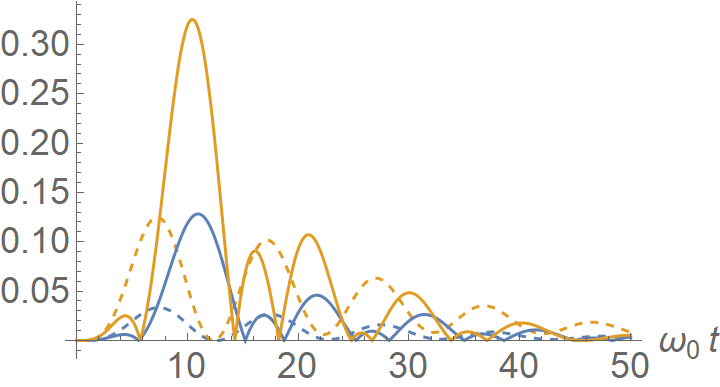}
    \caption{ }
    \label{fig:2CvCdiff}
\end{subfigure}  
\begin{subfigure}[b]{0.3\textwidth}
    \centering
    \includegraphics[width=\textwidth]{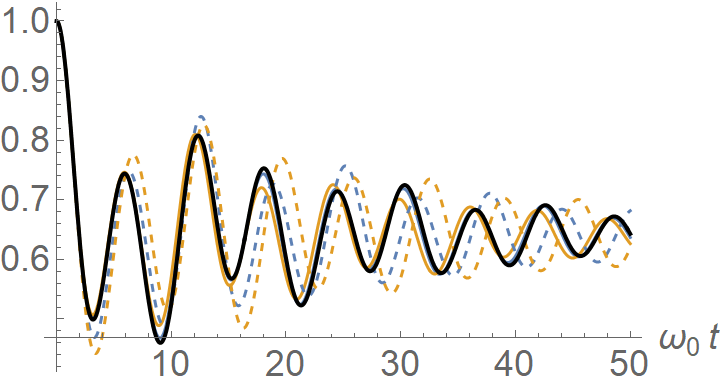}
    \caption{ }
    \label{fig:5A2CvC}
\end{subfigure}
\\
    
\begin{subfigure}[b]{0.3\textwidth}
    \centering
    \includegraphics[width=\textwidth]{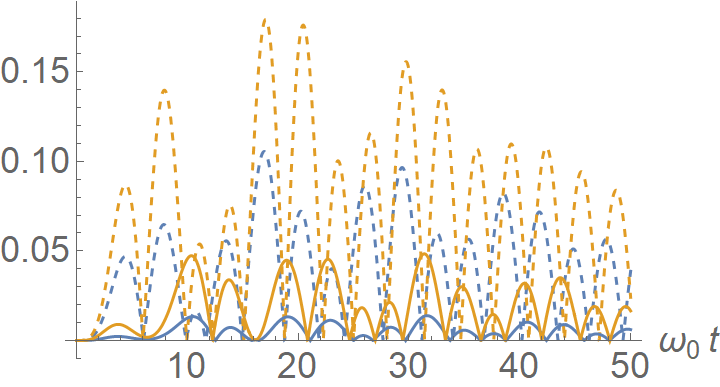}
    \caption{ }
    \label{fig:5A2CvCdiff}
\end{subfigure}
\begin{subfigure}[b]{0.25\textwidth}
    \centering
    \includegraphics[height=\textwidth]{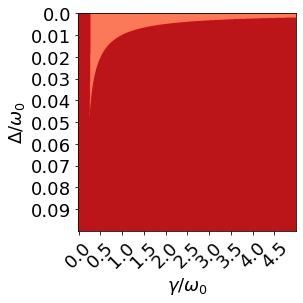}
    \caption{ }
    \label{fig:NMl0.01}
\end{subfigure}
\begin{subfigure}[b]{0.25\textwidth}
    \centering
    \includegraphics[height=\textwidth]{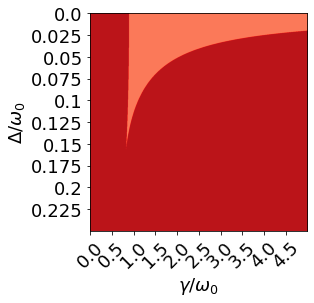}
    \caption{ }
    \label{fig:NMl0.1}
\end{subfigure}
    \caption{Plots of the time evolution of the excited state population of a single atom in a (thick) double (dashed) single detuned  (black) on resonance Lorentzian cavity. (a) For a single atomic system. (b) The magnitude of the difference between the plots in (a) and the resonant system. (c)  Coupled with four atomic systems. (d) The magnitude of the difference between the plots in (c) and the resonant system. For the two-cavity system $\lambda=0.05\omega_0$, for the single cavities $\lambda = 0.1\omega_0$, $\gamma =0.1\omega_0$ and $\Lambda=0.1\omega_0$. (e),(f) Heatmaps of the Cardano discriminant showing the region wherein the single atom system coupled to the double symmetrically detuned cavities undergoes oscillatory dynamics (light) no oscillations (dark) oscillations occur for $\lambda = $(e)$0.01\omega$ (f) $0.1\omega_0$. }
    \label{fig:2Call}
\end{figure*}
For a pair of cavities detuned from the transition energy of the atomic system by a given value $\pm\Delta$, the spectral density associated is given by 
\begin{equation}
    J(\omega) = \frac{\lambda\gamma}{(\omega-(\omega_0+\Delta))^2+\gamma^2} + \frac{\lambda\gamma}{(\omega-(\omega_0-\Delta))^2+\gamma^2},
\end{equation}
which leads to the associated memory kernel
\begin{equation}
    G(t) = 2\lambda e^{-\gamma t}\cos{\Delta t}.
\end{equation}

In the following, we are analysing the excited state populations obtained by solving the corresponding equations of motion given in Eqn.~\ref{eqn:EOM}. In the Fig.~\ref{fig:2Call}(a)-(d), we are comparing the case of the two cavities with symmetric detuning to the on-resonant case $\Delta=0$ and the single off-resonant cavity with twice the coupling strength. In Fig.~\ref{fig:2CvC}, we consider the time evolution of a single atomic system coupled to these three different cavity configurations. The double cavity system, although of the same coupling strength, leads to more rapid dissipation of the atomic excitation when compared to the on-resonant case. It further shows the inverse relation to the single cavity system, wherein the greater dephasing induces more rapid dissipation. Conversely, in Fig.~\ref{fig:5A2CvC}, wherein we have five coupled atomic systems, the double cavity system appears to be much more like the on-resonant case. This is further exemplified in Fig.~\ref{fig:5A2CvCdiff}, where we have plotted the relative difference between these configurations and the resonant cavity case. We can see that the two-cavity system across time remains much closer to the resonant case compared to the single on-resonant cavity. This suggests that if detuned cavities are inevitable in a physical platform, we could engineer this dephasing to make our system appear to act on resonance. 

In Appendix.\ref{app:doubcav} we have shown the solution to the differential equation in the Laplace transform frame. Via the inverse Laplace transform, we can infer that the rates in our dynamics will be given by the roots of the polynomial given by Eqn.\ref{eqn:DCPoly}.
From the Cardano formula, we can identify the regime under which the system undergoes non-Markovian dynamics. This is done by determining when these rates are real or complex, leading to Markovian and non-Markovian dynamics respectively. Complex valued rates will lead to oscillations in the population dynamics of the single atom. The discriminant is given by 
\begin{equation}
    \Delta_0 = \left(3 \Delta^2-\gamma^2+6 \lambda\right)^3+\left(9 \Delta^2 \gamma+\gamma^3-9 \gamma \lambda\right)^2,
\end{equation}
when $\Delta_0$ is negative, the system undergoes Markovian dynamics, and for positive values, the dynamics are non-Markovian. Heatmaps of this result are shown in Fig.~\ref{fig:NMl0.01} and~\ref{fig:NMl0.1}. Remarkably, we find that as we expand the non-Markovian regime for greater values of the detuning this enables the atomic system to regain the excitation from the reservoir By increasing the coupling strength, we can again increase this region of non-Markovianity. 

\section{Hierarchical Equations of Motion}
\begin{figure*}[ht!]
    \centering
     \begin{subfigure}[b]{0.3\textwidth}
    \centering
    \includegraphics[height=\textwidth]{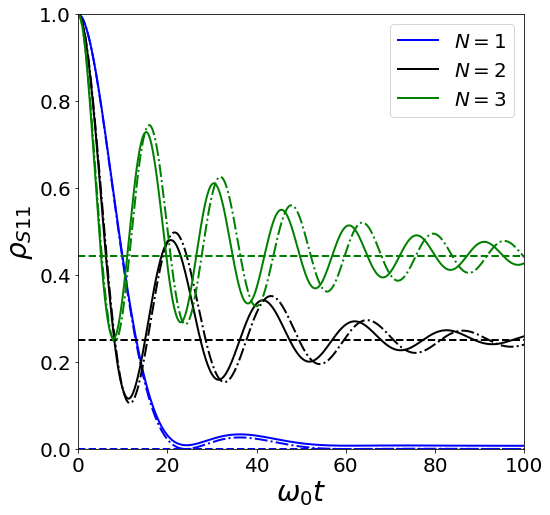}
    \caption{ }
    \label{fig:HEOMweak}
\end{subfigure}
\hfill
\begin{subfigure}[b]{0.3\textwidth}
    \centering
    \includegraphics[height=\textwidth]{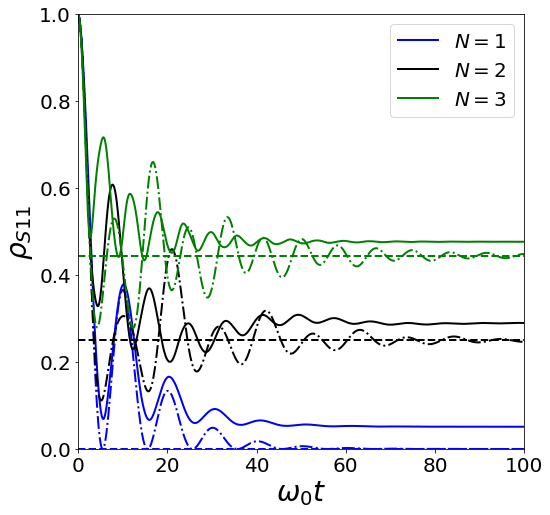}
    \caption{ }
    \label{fig:HEOMstrong}
\end{subfigure}  
\hfill
\begin{subfigure}[b]{0.3\textwidth}
    \centering
    \includegraphics[height=\textwidth]{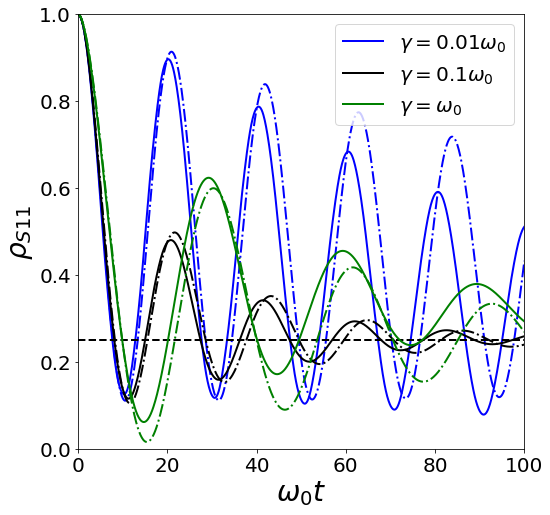}
    \caption{ }
    \label{fig:HEOMvarg}
\end{subfigure}
\\
\begin{subfigure}[b]{0.3\textwidth}
    \centering
    \includegraphics[height=\textwidth]{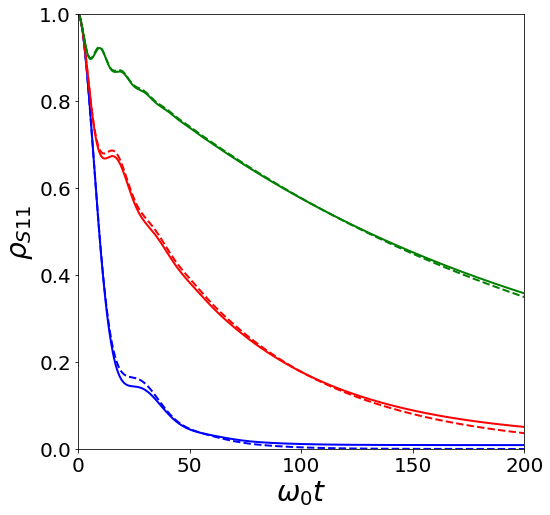}
    \caption{ }
    \label{fig:HEOMdetune}
\end{subfigure}
\hfill
\begin{subfigure}[b]{0.3\textwidth}
    \centering
    \includegraphics[height=\textwidth]{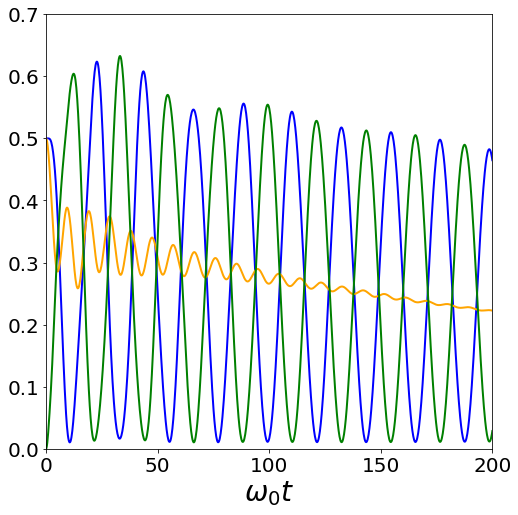}
    \caption{ }
    \label{fig:HEOMchain}
\end{subfigure}
\hfill
\begin{subfigure}[b]{0.3\textwidth}
    \centering
    \includegraphics[height=\textwidth]{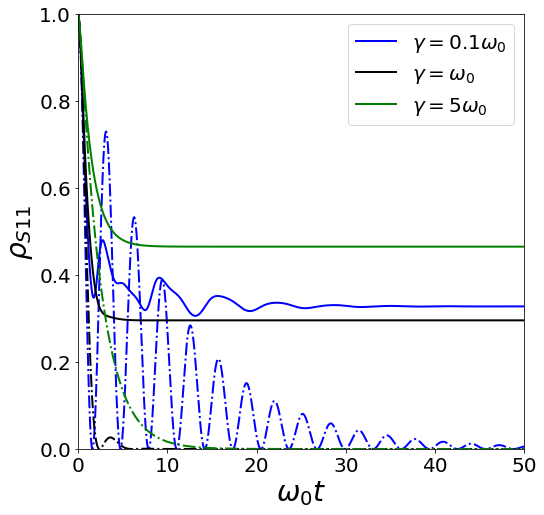}
    \caption{ }
    \label{fig:HEOMvstrong}
\end{subfigure}

    \caption{Plots of the time evolution of the excited state population of atoms in an on-resonance Lorentzian cavity solved using HEOM (thick) and RWA (dashed/dashed dot). (a) For the first atom initially excited for varying numbers of atoms in the system, with all to all coupling in the weak coupling regime $\lambda =0.01\omega_0$ with RWA predicted steady states (dashed). (b) as in (a) in the intermediate coupling regime $\lambda = 0.1\omega_0$. (c)  Two coupled atoms for varying spectral widths $\gamma$.
    (d) HEOM (thick) vs RWA (dashed) predictions for a single atom inside of a detuned cavity with detuning $\Delta =0.1\omega_0$(blue), $0.3\omega_0$ (red) and $0.6\omega_0$ (green). (e) HEOM results for Three atom chain system, with initial conditions $c_1(0)=-c_2(0)=\frac{1}{\sqrt{2}}$ for the population of the first (blue) second (orange) third (green) atom in the chain. (f) HEOM results for a single atom coupled strongly ($\lambda = \omega_0$) to the cavity for a variety of spectral widths $\gamma$. For all plots unless specified $\lambda = 0.01\omega_0$, $J=0.1\omega_0$, $\gamma=0.1\omega_0$. }
    \label{fig:HEOM}
\end{figure*}
Throughout this paper, we have deployed the rotating wave approximation. This is a common approximation within the field of quantum optics that greatly reduces the complexity of physical problems by neglecting rapidly oscillating terms in the Hamiltonian, allowing for the restriction of the Hilbert space to a fixed excitation number. For the systems considered in this article (atomic systems with equal transition energies and Lorentzian photonic cavities on resonance), this approximation is appropriate as most frequencies are approximately equal to the transition energy $\omega_0$. Therefore, most counter-rotating terms will oscillate with frequencies $2\omega_0$ and they can be effectively neglected.

In this section, we explore the accuracy of the RWA and validate our results by deploying the hierarchical equations of motion (HEOM)\cite{HEOM}. These are a set of differential equations capable of modelling the dynamics of a quantum system linearly coupled to a bosonic environment. This is precisely the configuration considered here. Furthermore, these equations of motion are numerically exact, making them non-perturbative and non-Markovian, imposing no limiting approximations on their solutions. The HEOM approach utilises a discretisation of the environment to generate a numerically efficient approach to capturing the effects of the collective degrees of freedom of the environment. This is achieved by the construction of coupled differential equations that form a hierarchy from repeated time-derivatives of the influence functional, derived initially by Vernon and Feynman\cite{FEYNMAN1963118}. However, with this approach come some shortcomings. Firstly, it is a numerical approach and will not lead to analytical results making the search of the parameter space less informed and less efficient. Secondly, HEOM requires that the bath correlation function is of an exponential form. Generally, this is the case for a limited number of spectral densities, but, fortunately, the Lorentzian photonic cavity spectral density considered does take this form. The bath correlation function at zero temperature is given by the Fourier transform of the spectral density.
\begin{equation}
    C(t) = \int^\infty_0 d\omega J(\omega) e^{-i\omega t},
\end{equation}
and the Lorentzian cavity system, this simply yields
\begin{equation}
    C(t) = \lambda e^{-(\gamma+i\omega_0)t}.
\end{equation}
We can then rewrite this as an exponential sum over real and imaginary parts 
\begin{gather}
    C(t) = C_R(t) +iC_I(t),\\
    C_R(t) = \frac{\lambda}{2}(e^{-(\gamma  + i \omega_0)t}+e^{-(\gamma  - i \omega_0)t}= \sum_{k=1}^2c_k^Re^{-\gamma_k^R},\\
    C_I(t) = \frac{\lambda}{2i}(e^{-(\gamma  + i \omega_0)t}-e^{-(\gamma  - i \omega_0)t} = \sum_{k=1}^2c_k^Ie^{-\gamma_k^I}.
\end{gather}
In the non-rotating wave approximation, we can divide the total Hamiltonian into the system Hamiltonian, containing only atomic degrees of freedom and the interaction Hamiltonian, as follows
\begin{align}
    H = H_S+H_I,
\end{align}
with the system Hamiltonian
\begin{equation}
   H_S = \sum_{i=1}^N\omega_0\sigma_i^+\sigma_i^- +\sum_{i< j}\Lambda\sigma^x_i\sigma_j^x,
\end{equation}
and the interaction Hamiltonian
\begin{gather}
    H_I = i\sum_{i,\lambda}\sigma_i^xg_\lambda(a_\lambda -a^\dag_\lambda )\\
    = i\sigma_T^x \sum_\lambda g_\lambda(a_\lambda -a^\dag_\lambda ),
\end{gather}
and $\sigma_T^x = \sum_{i=1}^N \sigma_i^x$.

The hierarchical equations of motion are then\cite{HEOM} 
\begin{align}
    \Dot{\rho}^n(t) &= \left( -iH^\times_S-\sum_{j=R,I}\sum_{k=1}^{2}n_{jk}\gamma^j_k \right)\rho^n(t) \nonumber \\ &-i\sum^{2}_{k=1}c_k^R n_{Rk}\sigma_x^{T\times} \rho^{n^-_{Rk}}(t) + \sum^{2}_{k=1}c_k^In_{Ik}\sigma_x^{To}\rho^{n^-_{Ik}}(t) \nonumber \\
    &-i\sum_{j=R,I}\sum_{k=1}^{2}\sigma_x^{T\times} \rho^{n^+_{jk}}(t),
    \label{eqns:heom}
\end{align}
where we used the multi-index $n = (n_{R1},n_{R2},n_{I1},n_{I2}), n_{R,Ii}\in \{0,...,N_c\}$ with $N_c$  the cut-off parameter defining the depth of the hierarchy needed for a converged result. We have also introduced the commutator and anticommutator notation $\sigma_x^{T\times} = [\sigma_x^T,\cdot]$ and $\sigma_x^{To} = \{\sigma_x^T,\cdot\}$. 
In Eqn.~\ref{eqns:heom}, the only physical density matrix is the $(0,0,0,0)$ indexed matrix which is the reduced atomic systems' density matrix. All other $\rho^n$ are auxiliary and encapsulate the environmental attributes. The density operators with index $n^\pm_{jk}$ refer to auxiliary density operators with the $k$th index of the real or imaginary term raised or lowered by one. For example, for index
\begin{align}
    n &= (1,2,1,3)\\
    n^+_{R2} &= (1,3,1,3).\nonumber
\end{align}
In Fig.~\ref{fig:HEOM}, we compare the RWA solutions to the numerically exact HEOM solutions. We note that the RWA begins to break down as we increase the coupling strength of the system, limiting the utility of this approximation to weak coupling regime. In Fig.~\ref{fig:HEOMweak} we consider a coupling strength of $\lambda=0.01\omega_0$ and we note a good agreement between both the HEOM and RWA predictions. However, at later  times, a phase difference between HEOM and RWA emerges as expected since the counter-rotating terms become more relevant. This has little effect on the steady state characteristics and the RWA and HEOM predict nearly identical steady states for the system. This is quite remarkable as in the RWA we restrict our dynamics to a greatly reduced subspace of the entire Hilbert space. The total Hamiltonian does not conserve excitation number and the dynamics can traverse larger domains of the Hilbert space, yet we still have the RWA predicted steady state. 
Increasing the coupling strength in Fig.~\ref{fig:HEOMstrong}, we note  that the RWA approximation starts to break down, and the agreement between the RWA and the HEOM approaches is only qualitative. 
This is due to the discrepancy between the RWA Hamiltonians and the total Hamiltonian, the system's ground state is not equivalent. In RWA, the ground state is associated with zero excitations in the entire system. In the total Hamiltonian description, we have the ground state associated with mixed states of the reservoir degrees of freedom and the atomic systems. The degree of mixing is proportional to the coupling strength of the atomic systems to the reservoir $\lambda$. Therefore, as we increase the coupling strength, we have greater mixing in the total system steady state, and we see a more significant divergence between the RWA and HEOM steady states.

Furthermore, as the coupling strength increases, we increase the number of non-excitation number conserving interactions, leading to further deviation. In Fig.~\ref{fig:HEOMvarg}, we have varied the spectral width of the cavity and compared the HEOM and RWA solutions. Here we have two conflicting effects, as we decrease spectral width, we increase the coupling strength to the central cavity mode, making the RWA less reliable. However, by reducing the width, we also reduce the effect of non-resonant modes in the cavity, which should improve the accuracy of RWA. As such, we note that the intermediate value of the spectral width $\gamma=0.1\omega_0$ provides the best agreement.
In Fig.~\ref{fig:HEOMdetune}, somewhat surprisingly, we find that in the weak coupling regime, the HEOM and the RWA approximation results coincide very well for detuned cavities, even across large values of the detuning, suggesting that we can learn about detuned systems within the RWA approximation.
Remarkably, we are able to validate the phenomenon of reduced dissipation within the atomic chain which leads to almost free dynamics for the two ends of the chain in Fig.~\ref{fig:HEOMchain}, confirming that is not an artefact of the RWA.

Due to the non-perturbative nature of HEOM we can also explore the strong coupling regime. In Fig.\ref{fig:HEOMvstrong} we consider the dynamics of a single atom strongly coupled ($\lambda = \omega_0$) to the cavity for a few parameters of the spectral width ($\gamma$). Here we can see additional phenomena emerging. Firstly, we can see that the for the case wherein $\gamma=\omega_0$ we have exponential decay in HEOM, however, within the RWA, we expect to see oscillatory dynamics. This shows that the strong coupling limit begins to modulate the condition on oscillatory dynamics away from the standard $\gamma^2<4\lambda$ relation. Furthermore, we note that for the narrow spectral width $\gamma =0.1\omega_0$, we have an introduction of at least one additional phase into the oscillatory dynamics moving away from a single phase enveloped by a decay. Interestingly, we do see that the RWA predictions does seem to coincide with the peaks of the HEOM calculations, showing that some of the RWA dynamics remain, the RWA predictions also give good qualitative agreement of the decay rates with HEOM. Finally, we find that the largest spectral width $\gamma=5\omega_0$ has the greatest discrepancy from the HEOM, this is particularly startling as the effective coupling strength here is lower for higher spectral width, yet we see that this configuration has the highest overall steady state.

As we contrast the RWA and HEOM approaches, we note that in the weak and intermediate coupling regime, the RWA analytical formalism provides valuable insights into the actual physical dynamics of the system and can guide our search using numerically exact techniques. On the other hand, HEOM provides us with a powerful tool to study parameter regimes inaccessible in RWA.

\section{Conclusion}
To conclude, we have explored the dynamics of many-body atomic systems symmetrically coupled to a single Lorentzian photonic cavity and identified a number of intriguing dynamical traits. This includes superradiant decay, non-zero stable states, improved energy transfer, and the capacity to control oscillations in the atomic system by adjusting environmental degrees of freedom. We also explored multi-atomic systems coupled via  nearest neighbour interaction to each other and to the photonic reservoir of modes of the cavity. In this case, we predict a similar increase in the rate of energy transfer between the two ends of the chain. We have also identified the initial conditions which result in a significant reduction of and exchange excitations between the two atoms at the end of the chain. We further demonstrated the ability of the structured photonic reservoir to influence the atomic evolution in the case of two symmetrical detuned reservoirs. Here, for the single-atom case, we show that is possible to increase the system's decay rate by modulating its detuning with respect to the reservoir spectral features. In the many-atom case, we show that reservoir structuring brings the system dynamics' closer to the on-resonant cavity case. Finally, we demonstrated the RWA predictions by deploying a numerically exact HEOM approach and extended our results in the regime in which the RWA is no longer valid. Our results demonstrate that despite the dissipative nature of the photonic reservoirs,  careful choices of initial conditions or engineering of the environment can provide a robust control over the embedded atomic systems dynamics.

\begin{acknowledgments}
This work was supported by the Leverhulme Quantum Biology Doctoral Training Centre at the University of Surrey were funded by a Leverhulme Trust training centre grant number DS-2017-079, and the EPSRC (United Kingdom) Strategic Equipment Grant No. EP/L02263X/1 (EP/M008576/1) and EPSRC (United Kingdom) Grant EP/M027791/1 awards to M.F. We acknowledge helpful discussions with the members of the Leverhulme Quantum Biology Doctoral Training Centre.
\end{acknowledgments}

\appendix

\section{Single Reservoir}
\subsection{Fully Symmetric Coupling}
\label{app:FC}
We start with the (RWA) Hamiltonian 
\begin{align}
    H =& \sum_{i=1}^N\omega_0\sigma_i^+\sigma_i^- + \sum_\lambda \omega_\lambda a^\dag_\lambda a_\lambda \nonumber\\&+ i \sum_{i,\lambda}g_\lambda(a_\lambda \sigma^+_i -a^\dag_\lambda \sigma^-_i ) + \sum_{i\neq j}\Lambda\sigma^+_i\sigma_j^-,
\end{align}
where $\sigma_i^+$and $\sigma_i^-$, are the excitation and de-excitation operators for the $i^\textrm{th}$ atomic system respectively, $a_{\lambda}$ and $a^{\dagger}_{\lambda}$ are the bosonic field annihilation and creation operators, $\omega_0$ and $\omega_{\lambda}$ are the atomic transition and the $\lambda$-boson mode frequencies, and $g_\lambda$ is the coupling strength of the atomic system and the $\lambda$-boson mode.

By changing to the interacting picture, we can consider the interaction Hamiltonian 
\begin{align}
    \Tilde{H}_I =& i\sum_{i,\lambda}g_\lambda(a_\lambda \sigma^+_i e^{i(\omega_0-\omega_\lambda)t} -a^\dag_\lambda \sigma^-_i e^{-i(\omega_0-\omega_\lambda)t} )\nonumber \\& + \sum_{j\neq k}\Lambda\sigma^+_j\sigma_k^-.
\end{align}
Such a Hamiltonian is convenient as it conserves the excitation number of any wavefunction. As such, if we consider the single excitation wavefunction given by 
\begin{equation}
    \phi(0) = c_0\psi_0+\sum_i^N c_i(0)\psi_i +\sum_\lambda c_\lambda(0) \psi_\lambda,
\end{equation}
its time evolution is given by 
\begin{equation}
    \phi(t) = c_0\psi_0 +\sum_i^N c_i(t)\psi_i +\sum_\lambda c_\lambda(t) \psi_\lambda,
\end{equation}
where $\psi_i = \ket{i}_A\ket{0}_B$ the ith atom is in its excited state and $\psi_\lambda = \ket{0}_A\ket{\lambda}_B$ the atoms are all in their ground state, and the bosonic system has it is $\lambda$ mode excited. By simply plugging this into the Schr{\"o}dinger equation and determining the coupled differential equations of motion for the state amplitudes, we get 
\begin{equation}
    \Tilde{H}_I \phi_i = -i\sum_\lambda g_\lambda \psi_\lambda e^{-i(\omega_0-\omega_\lambda)t} + \sum_j^N \Lambda \psi_j (1- \delta_{ij}),
\end{equation}

\begin{equation}
    \Tilde{H}_I \phi_\lambda = ig_\lambda e^{i(\omega_0-\omega_\lambda)t} \sum_i \psi_i.
\end{equation}

By introducing the parameter 
\begin{equation}
    c_+(t) = \sum_i c_i(t)  , 
\end{equation}
we have 
\begin{align}
    \Dot{c}_i =& -i\Lambda(c_+-c_i) + \sum_\lambda c_\lambda g_\lambda e^{i(\omega_0-\omega_\lambda)t},\nonumber \\
    \Dot{c}_+ =& -i\Lambda(N-1)c_+ + N \sum_\lambda c_\lambda g_\lambda e^{i(\omega_0-\omega_\lambda)t} ,\nonumber \\
    \Dot{c}_\lambda =& -g_\lambda e^{-i(\omega_0-\omega_\lambda)t} c_+.
\end{align}
Assuming the bosonic field is initially in its vacuum configuration ($c_\lambda(0) = 0$), we can formally integrate up the last equation to get 
\begin{equation}
    c_\lambda(t) = -\int^t_0 g_\lambda e^{-(\omega_0-\omega_\lambda)t_1}c_+(t_1)dt_1,
\end{equation}
and by introducing the memory kernel $G(t) = \sum_\lambda g_\lambda^2 e^{i(\omega_0-\omega_\lambda)t}$ we can rewrite the above equations as 

\begin{align}
    \Dot{c}_i =& -i\Lambda(c_+-c_i) - \int^t_0G(t-t_1)c_+(t_1)dt_1,\nonumber \\
    \Dot{c}_+ =& -i\Lambda(N-1)c_+ - N\int^t_0G(t-t_1)c_+(t_1)dt_1.    
\end{align}
It is notable that our memory kernel $G(t)$ is related to the spectral density $S(\omega)= \sum_\lambda g_\lambda^2 \delta(\omega-\omega_\lambda)$ associated to the reservoir by the relation
\begin{equation}
G(t) = \int d\omega S(\omega )e^{i(\omega_0-\omega)t}.
\end{equation}
In order to solve these equations, we utilise the Laplace transform. Transforming the differential equations into algebraic ones.

\begin{align}
    s\Tilde{c}_i(s) - c_i(0) &= -i\Lambda(\Tilde{c}_+(s) -\Tilde{c}_i(s)) - \Tilde{c}_+(s)\Tilde{G}_1\nonumber, \\
    s\Tilde{c}_+(s) - c_+(0) &= -i\Lambda(N-1)\Tilde{c}_+(s) - N\Tilde{c}_+(s)\Tilde{G}_1.
\end{align}
Where we have used that $\Tilde{f}(s)=\mathcal{L}\{f(t)\}$. We can see that we can solve for $\Tilde{c}_+(s)$ getting

\begin{equation}
    \Tilde{c}_+(s) = \frac{c_+(0)}{s+i\Lambda(N-1)+N\Tilde{G}}.
\end{equation}
From this, we can solve for the single atom amplitude Laplace solution 
\begin{equation}
    \Tilde{c}_i(s) = \frac{c_i(0)}{s-i\Lambda} - \frac{c_+(0)(\Tilde{G}+i\Lambda)}{(s-i\Lambda)(s+i\Lambda(N-1)+N\Tilde{G})}.
\end{equation}
If we consider the form of $\Tilde{G}(s)$ with respect to the spectral density, we can see that

\begin{equation}
\Tilde{G}(s) =\int d\omega \frac{S(\omega)}{s-i(\omega_0-\omega)}.
\end{equation}

\section{Double Cavity System}
\label{app:doubcav}
The Laplace transform of the memory kernel is given by 
\begin{equation}
    \Tilde{G}(s) = \frac{2\lambda(\gamma+s)}{\Delta^2 + (\gamma +s)^2}.
\end{equation}

In Laplace space the solutions to the equations of motion are given by 
\begin{gather}
    c_i(s) = \frac{c_i(0)}{s-i\Lambda}-\\
    \frac{c_+(0)\left(i\Lambda(\Delta^2+(\gamma+s)^2)+2\lambda(\gamma+s)\right)}{(s-i\Lambda)\big((i\Lambda(N-1)+s)(\Delta^2+(\gamma+s)^2)+2\lambda N(\gamma+s)\big)}.
\end{gather}

Considering just the single atom case $N=1$, $J=0$. We have that 
\begin{gather}
    c_i(s) = \frac{c_i(0)(\Delta^2+(\gamma+s)^2)}{s(\Delta^2+(\gamma+s)^2)+2\lambda(\gamma+s)}.
\end{gather}
The roots of the denominator will determine the rates of the equation by the inverse Laplace transform. The expanded denominator will give 
\begin{equation}
    \lambda \gamma + (\Delta^2  + \gamma^2  + 2 \lambda) s + 2 \gamma s^2 + s^3.
    \label{eqn:DCPoly}
\end{equation}

\bibliography{main}
\end{document}